\documentclass[aps,prc,twocolumn,amsmath,amssymb,nofootinbib]{revtex4-2}

\usepackage{graphicx}
\usepackage{tikz}
\usepackage{xcolor}
\definecolor{reviewpurple}{RGB}{110,55,145}
\definecolor{brown}{RGB}{140,80,20}
\usepackage[hidelinks]{hyperref}

\def\be{\begin{eqnarray}}
\def\ee{\end{eqnarray}}

\newcommand{\nucst}[3]{{}^{#1}\text{#2}_{#3}}

\def\vsigma{\vec \sigma}
\def\vp{\vec p}
\def\vq{\vec q}
\def\vk{\vec k}
\def\vtau{\vec \tau}
\def\vpi{\vec \pi}

\def\tt{\tau_1\cdot\tau_2}
\def\ssig{\vsigma_1\cdot\vsigma_2}
\def\sqq{\vsigma_1\cdot \vq\, \vsigma_2\cdot \vq}
\def\no{\nonumber}

\def\OLS{{\cal O}_{LS}}

\begin{document}

\title{Peripheral nucleon--nucleon scattering with explicit vector mesons
in chiral perturbation theory}

\author{Tae-Sun Park}
\email{tspark@ibs.re.kr}
\affiliation{Center for Exotic Nuclear Studies, Institute for Basic Science, Daejeon 34126, Korea}

\date{\today}

\begin{abstract}
\begin{description}

\item[Background] In chiral perturbation theory (ChPT) the nucleon--nucleon interaction is built from pions and nucleons,
and other heavy degrees of freedom are integrated out.
Their contributions are embodied in the low-energy constants of the operators with increasing numbers of derivatives.
Keeping the $\rho$ and $\omega$ explicit within ChPT retains the finite-range momentum dependence associated with these mesons.

\item[Purpose] We study how retaining the $\rho$ and $\omega$ explicitly within ChPT affects the peripheral $NN$ phase shifts.

\item[Methods] We incorporate the vector mesons by counting their masses as hard and the momentum they carry as soft.
The next-to-next-to-leading-order nuclear force includes,
in addition to one- and two-pion exchange,
one-vector-meson exchange with dressed propagators and vertex functions.
The vector propagators and vertex functions are fixed from the mesonic sector and the nucleon electromagnetic form factors. 
We calculate neutron--proton scattering with $J=3$--$5$ up to $T_{\rm lab}=300$~MeV and examine the sensitivity to one-pion-exchange iteration and the spectral-function regularization scheme.

\item[Results] The vector-meson contribution falls rapidly with $J$.
In several $J=3$ waves it moves the phase shifts toward PWA93,
while in other waves the change is small or moves the phase shift away.
In ${}^{3}\mathrm{G}_{3}$ the change toward PWA93 persists under variations of the spectral cutoff,
and its size is nearly independent of the treatment of one-pion exchange.
The ${}^{3}\mathrm{D}_{3}$ wave is strongly cutoff dependent,
and there the $\omega$ contribution outweighs the opposing $\rho$ contribution.
The tensor and spin-spin components of the $\rho$ interaction play the dominant roles in moving $\epsilon_3$ and ${}^{1}\mathrm{F}_{3}$ toward PWA93, respectively.

\item[Conclusions] The effect of the explicit vector mesons on the peripheral phase shifts is channel dependent.
The $\rho$ meson gives the dominant shifts toward PWA93,
particularly in $\epsilon_3$ and ${}^{3}\mathrm{G}_{3}$.

\end{description}
\end{abstract}


\maketitle

\section{Introduction\label{sec-intro}}

Chiral perturbation theory (ChPT) organizes nuclear interactions as an
expansion in a soft scale $Q$ over a breakdown scale of order
$\Lambda_\chi\sim1$~GeV.  In the conventional ChPT formulation, pions
and nucleons are retained explicitly, while heavier degrees of freedom are integrated out
~\cite{weinberg,EHM-review,machleidt-entem}.  An alternative organization is
to keep selected resonances explicitly.  
Treating them in this way reorganizes their contributions and retains the momentum dependence that would otherwise be absorbed into the low-energy constants (LECs) of higher-order operators.
Similar reorganizations arise in formulations with an explicit $\Delta(1232)$~\cite{krebs2007,jiang2020,nosyk2021} or Roper resonance~\cite{xgvk2026}.

The $\rho$ and $\omega$ are natural candidates for such a reorganization of the $NN$ interaction.
They have long played an important role in the nuclear force,
in particular by providing short-range repulsion.
In addition, the $\rho$ produces a substantial cancellation of the one-pion-exchange tensor interaction~\cite{kaiser-meissner1990,rho2016}. 
These features motivate treating the $\rho$ and $\omega$ explicitly within ChPT and examining how their inclusion modifies the $NN$ interaction.

For this purpose, we adopt the counting of Refs.~\cite{pmr,rho91,paper1,fsgs2003},
in which the vector-meson masses are treated as hard,
$m_V\sim\Lambda_\chi$, and the momentum transfer as soft, $q\sim Q$.
A key feature of this counting is that the vector-meson kinetic term is suppressed by $Q^2/\Lambda_\chi^2$ relative to the mass term,
and does not enter the leading-order (LO) Lagrangian.
The leading vector-meson propagator is therefore momentum independent,
so its contribution can be absorbed into the leading $NN$ contact interactions.
For the same reason, loops with the leading-order vector-meson propagators reduce to those of ChPT. 
Resolved finite-range effects enter at higher orders through the self-energy and vertex corrections of the vector mesons.

Peripheral partial waves 
are
free of the LECs of
contact interactions, which enter a partial wave of orbital angular momentum
$L$ first at order $Q^{2L}$, and they have been used extensively to test
pion-exchange dynamics in chiral effective field theory~\cite{kbw,ekmn2015}.
In particular,
Kaiser, Gerstend\"orfer and Weise~\cite{kgw1998} studied the effect of
phenomenological one-$\rho$- and one-$\omega$-exchange added to the chiral
pion-exchange interaction, and found sizable vector-meson effects in the $F$
waves and a smaller contribution to $\epsilon_3$.
The distinguishing feature of the present work is the systematic inclusion of vector mesons in ChPT with a definite power counting.
The LECs appearing at each order are fixed by the same low-energy data as in ChPT, so that for
momenta well below the vector-meson masses the potential agrees with the ChPT
potential order by order.
Among the subleading $\pi N$ LECs, only $c_4$ is affected,
and it is shared between pion exchange and the explicit $\rho$.
The $\rho$ carries the
part $\kappa_\rho/(4m_N)$~\cite{bernard1997},
and $c_4$ is renormalized in such a way as to retain the low-energy $\pi N$
dynamics.  

Two-pion exchange is regularized with a spectral cutoff~\cite{egm-sfr},
which removes its spectral strength above a mass $\tilde\Lambda$.  With an
explicit $\rho$ this requires care.  The $\rho$ self-energy contains the same
two-pion states, and the LECs of the $\rho$ sector are fixed on the $\rho$
mass shell~\cite{paper1}.  A cutoff below the $\rho$ mass would remove the
$\rho\to\pi\pi$ decay, which complicates the on-shell renormalization of the
$\rho$ sector.  We therefore consider two cases (Sec.~\ref{sec-reg}).  In the
first, the cutoff is applied to two-pion exchange alone.
In the second, a cutoff above the $\rho$ mass is applied to both two-pion exchange and the $\rho$ sector.
For reference, we also show the results in the unregularized limit $\tilde\Lambda\to\infty$.
When the cutoff is applied to the $\rho$ sector, its LECs are refitted.

The vector-meson effects decrease rapidly with increasing $J$, and we study
the peripheral waves with $J=3$--$5$ at laboratory energies up to 300~MeV.
Other explicit resonances, in particular the $\Delta(1232)$, are beyond the scope
of this work.

Section~\ref{sec-vminput} describes ChPT with explicit vector mesons,
including the power counting, the Lagrangian and the dressed propagators.
Section~\ref{sec-force} constructs the two-nucleon interaction with explicit
vector mesons up to next-to-next-to-leading order (N$^2$LO).
In Sec.~\ref{sec-results} we apply it to the peripheral
waves, compare the phase shifts with the Nijmegen partial-wave analysis PWA93~\cite{nijmegen},
and examine their dependence on the iteration of one-pion exchange (OPE),
the spectral cutoff, the treatment of the vertex loops and the values of $\kappa_\rho$,
$\kappa_\omega$, $g_\omega$ and $c_4^{\rm VM}$.
The results are summarized in the last
section.

\section{Vector mesons in ChPT\label{sec-vminput}}

\subsection{Power counting and Lagrangian}

The mesonic Lagrangian is that of Ref.~\cite{paper1}.  We organize
${\cal L}={\cal L}_0+{\cal L}_1+{\cal L}_2+\cdots$, where
${\cal L}_\nu$ contains vertices of index~\cite{rho91,pmr}
\be
\nu_i = d_i + \frac{n_i}{2} + h_i + e_i - 2 .
\label{counting}\ee
Here $d_i$ counts derivatives or powers of the pion mass, $n_i$ the number
of nucleon fields, $h_i$ the number of heavy-meson fields, and $e_i$ the
number of external fields. 
The essential point of this counting is that the vector-meson kinetic term, with $d_i=2$ and
$h_i=2$, has $\nu_i=2$ and is not part of ${\cal L}_0$.

Without the kinetic term the leading vector propagator is static,
$g^{\mu\nu}/m_V^2$, and the corresponding exchange reduces to local
operators, which are absorbed into the leading contact terms.  Inside a loop
a vector-meson propagator likewise reduces to a local vertex.
The kinetic term enters at next-to-leading order (NLO) as a contribution to the self-energy,
together with the two-pion loop.
Resolved momentum dependence enters through the dressed propagators and the
loop corrections to the vertex functions starting from NLO.

The dressed vector-meson propagator can be represented as an infinite
series of powers of the self-energy.
Truncating this series and the vertex corrections at a given order recovers conventional ChPT at that order.
In this work the self-energies and vertex functions are obtained to a
given order, and the propagators and vertex functions are used in full rather
than expanded and truncated.  This resummation in the propagator is the main
mechanism of the reorganization described in the Introduction.  The present
results therefore differ from those of conventional ChPT only at orders
beyond those considered explicitly.

The Lagrangian can also be decomposed according to the number of nucleon fields,
${\cal L}={\cal L}^{(0N)}+{\cal L}^{(2N)}+{\cal L}^{(4N)}+\cdots$.  The leading
mesonic Lagrangian is
\be
{\cal L}^{(0N)}_0 &=& f_\pi^2\,\text{tr}\!\left(
i\Delta_\mu\, i \Delta^\mu\right)
+ \frac{f_\pi^2}{4}\,\text{tr}\,\chi_+
\no\\ &&
+\, a f_\pi^2\,
\text{tr}\!\left(g\rho_\mu - i\Gamma_\mu\right)^2
+ \frac12 m_\omega^2 \omega_\mu \omega^\mu ,
\label{L0}\ee
where $\Gamma_\mu$ and $\Delta_\mu$ are defined by
\be
\Gamma_\mu + \Delta_\mu &=& \xi^\dagger\partial_\mu\xi ,
\no\\
\Gamma_\mu - \Delta_\mu &=& \xi\,\partial_\mu\xi^\dagger ,
\label{conn}\ee
with $\xi=\sqrt U=\exp(i\vtau\cdot\vpi/2f_\pi)$,
$\chi_+=m_\pi^2\,(U+U^\dagger)$ and $\rho_\mu=\tau^a\rho^a_\mu/2$.
The KSRF parameter $a$ gives the
tree-level relation~\cite{ks1966,rf1966,hls,paper1}
\be
m_\rho^2 \simeq a g^2 f_\pi^2 ,
\label{ksrf}\ee
whose one-loop form is discussed in
the following subsection.

With the vector mesons present, the nucleon covariant derivative
${\cal D}_\mu=\partial_\mu+\Gamma_\mu$
becomes~\cite{borasoy1996,motohiro2015,djukanovic04}
\be
{\cal D}_\mu = \partial_\mu + \Gamma_\mu
- \frac{a_N}{2}\left(\Gamma_\mu + i g\rho_\mu + i g_\omega \omega_\mu\right).
\label{Dcov}\ee
The coefficient $a_N$ is a low-energy constant, the counterpart of $a$
in the nucleon sector.
The one-nucleon Lagrangian 
reads
\be
{\cal L}^{(2N)} &=& \bar N\left[i\gamma^\mu\partial_\mu - m_N
+ g_A\,\gamma^\mu\gamma_5\,i\Delta_\mu\right] N
\no\\ &&
+\ i\,c_4^{\rm VM}\,\bar N\left[i\Delta_\mu,i\Delta_\nu\right]\sigma^{\mu\nu} N
\no\\ &&
+\ {\cal L}_{VNN}\ +\ \cdots ,
\label{LVM}\ee
with
\be
{\cal L}_{VNN} &=&
\frac{a_N}{2}\, g\,\bar N\left[\gamma^\mu \rho_\mu
+ \frac{\kappa_\rho}{4m_N}\sigma^{\mu\nu}\rho_{\mu\nu}\right]N
\no\\ &&
+\ (\rho \to \omega,\ g \to g_\omega,\ \kappa_\rho \to \kappa_\omega) ,
\label{LVNN}\ee
where $\rho_{\mu\nu}=\partial_\mu\rho_\nu-\partial_\nu\rho_\mu-ig[\rho_\mu,\rho_\nu]$ and the ellipsis denotes other higher-order terms.
Of ${\cal L}_1$ only the two operators that enter the discussion of the
vector mesons are shown, the Pauli couplings $\kappa_\rho$, $\kappa_\omega$ and the $c_4$ term.
The latter
is the $c_4$ operator of ChPT, 
and $c_4^{\rm VM}$ denotes its coefficient in the theory with explicit vector mesons.
When the $\rho$ is integrated out, $g\rho_\mu$ becomes $i\Gamma_\mu$, so that
$g\rho_{\mu\nu}\to i[D_\mu,D_\nu]=i[i\Delta_\mu,i\Delta_\nu]$, and the Pauli term in
Eq.~(\ref{LVNN}) produces the same operator structure as the $c_4$ term in
Eq.~(\ref{LVM}), with coefficient $\kappa_\rho/(4m_N)$.
The constant $c_4$ is therefore renormalized to
\be
c_4^{\rm VM} = c_4^{\rm ChPT} - \frac{a_N}{2}\frac{\kappa_\rho}{4m_N} ,
\label{c4split}\ee
where $c_4^{\rm ChPT}$ is the constant of conventional ChPT, written $c_4$ elsewhere in this paper.

The remaining constant $a_N$ could be adjusted in a precision potential.
Here we take
\be
a_N=a=2
\label{aN}\ee
throughout this work.  With this choice the photon reaches the nucleon only
through the vector mesons, as it reaches the pion only through the $\rho$ at
$a=2$~\cite{paper1}, and the $\rho NN$ coupling $g_{\rho NN}=(a_N/2)g$
equals the $\rho\pi\pi$ coupling $g$~\cite{djukanovic04}.
We also take the vector-meson-dominance values
\be
\kappa_\rho&=&\mu_p-\mu_n-1=3.706,
\no\\
\kappa_\omega&=&\mu_p+\mu_n-1=-0.120 ,
\label{kapvalues}\ee
as reference inputs.
With these values the vector-meson poles carry the full anomalous magnetic moments.

\subsection{Vector-meson propagators\label{sec-prop}}

At leading order the propagator is momentum independent,
$D_V^{\mu\nu(0)}(k)=g^{\mu\nu}/m_V^2$.
At NLO the $\rho$ propagator acquires its momentum dependence through the
kinetic term and the two-pion loop of the self-energy~\cite{paper1},
\be
D_\rho^{\mu\nu}(k) &=&
D_\rho(t)\,
\left(g^{\mu\nu}-\frac{\Pi_\rho(t)}{2g^2f_\pi^2}
\frac{k^\mu k^\nu}{t}\right),
\nonumber\\
D_\rho(t) &=& \frac{1}{2g^2f_\pi^2-\Pi_\rho(t)} ,
\label{Drho}\ee
with $t=k^2$ and
\be
\Pi_\rho(t) &=& t\,A_\rho(t) ,
\no\\
A_\rho(t) &=& A_\rho(0)
+2\left[\Gamma_{\rho\pi\pi}(t)\right]^2\bar f_3(t) ,
\label{Arho}\ee
where $\bar f_3(t)$ is the loop function of Eq.~(\ref{f3bar}),
with $\bar f_3(0)=0$, and $\Gamma_{\rho\pi\pi}(t)$ is the one-particle-irreducible $\rho\pi\pi$ vertex, defined through
$-i\epsilon_{eab}(q_a-q_b)^\mu\,\Gamma_{\rho\pi\pi}(t)$ for
$\rho^\mu_e(k)\to\pi_a(q_a)\pi_b(q_b)$,
\be
\Gamma_{\rho\pi\pi}(t)
= g+\frac{t}{m_\rho^2}\left[
\Gamma_{\rho\pi\pi}(m_\rho^2)
-g\right] .
\label{Grpp}\ee
From $\Gamma(\rho\to\pi\pi)=147.4$~MeV~\cite{pdg} together with the position and residue of the
propagator pole, Ref.~\cite{paper1} obtains
\be
\Gamma_{\rho\pi\pi}(m_\rho^2) &=& 5.942 ,
\no\\
g &=& 6.21 ,
\no\\
A_\rho(0) &=& 1.113 .
\label{rhoinput}\ee
with $m_\rho=0.77545$~GeV.\footnote{The $\rho$-sector inputs of Ref.~\cite{paper1} are obtained with $f_\pi=92.07$~MeV.}

The $\omega$ kinetic term enters at NLO as well, and the $\omega$ takes no
loop correction at this order,
\be
D^{\mu\nu}_\omega(k)&=&
D_\omega(t)\left(g^{\mu\nu}- \frac{k^\mu k^\nu}{m_\omega^2}\right),
\no\\
D_\omega(t)&=&\frac{1}{m_\omega^2-t} ,
\label{Domega}\ee
with $m_\omega=0.783$~GeV.
In this work the $\omega NN$ coupling $g_\omega=8.53$ is adopted.
It
follows from $g_\omega=m_\omega/(2F_\omega)$, with $F_\omega$ fixed by
$\Gamma(\omega\to e^+e^-)=4\pi\alpha^2F_\omega^2/(3m_\omega)=0.60$~keV~\cite{kkw1996,pdg}.
In both propagators the $k^\mu k^\nu$ terms contract into nucleon currents conserved between on-shell states and drop out,
so only $D_\rho(t)$ and $D_\omega(t)$
enter the potential.

\begin{figure*}
\centering
\includegraphics[width=\textwidth]{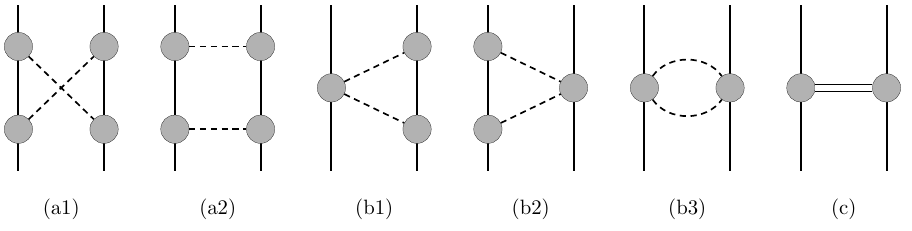}
\caption{Two-pion exchange, (a) and (b),
and one vector-meson exchange (c).
Solid, dashed and double lines denote
the full propagators of nucleons, pions and vector mesons, respectively.
Shaded circles are one-particle-irreducible vertices.
\label{fig-tpe}}
\end{figure*}

\section{The two-nucleon interaction\label{sec-force}}

At LO the two-nucleon potential consists of one-pion exchange
and contact operators without derivatives.  The leading part of
one-vector-meson exchange is also LO, but it is momentum independent and is
absorbed into these contact operators.  At NLO, the $Q^2$ correction, the
leading two-pion exchange (TPE) and contact operators with two derivatives
enter.  N$^2$LO, the $Q^3$ correction, adds the subleading TPE, the one-loop
graphs with one insertion of the $c_1$, $c_3$ and $c_4$ vertices of
${\cal L}_1$, and no new contact operator.
A contact operator with $n$ derivatives contributes to waves with $L\le n/2$,
so contact terms reach $\nucst{3}{D}{3}$ first at $Q^4$,
the $F$ waves at $Q^6$ and the $G$ waves at $Q^8$,
and the waves considered here are free of them at N$^2$LO.
Figure~\ref{fig-tpe} shows the
TPE graphs (a1)--(b3) and the one-vector-meson exchange, graph (c).
The ChPT potential up to this order is labelled N$^2$LO,
and N$^2$LO$_{\rm VM}$ denotes the potential of the same order with the explicit $\rho$ and $\omega$.

The one-pion exchange potential reads
\be
V_{1\pi}
=\frac{g_A^2}{4f_\pi^2}\left[
\frac{\sqq}{\vq^2+m_{\pi^0}^2}
+2 (-1)^{T}\,\frac{\sqq}{\vq^2+m_{\pi^\pm}^2}\right],
\label{VOPE}\ee
for the proton--neutron system in the isospin-$T$ channel,
with the momentum transfer $\vq=\vp\,'-\vp$, $g_A=1.29$
and $f_\pi=92.4$~MeV~\cite{machleidt-entem,rke2021}.

\subsection{One-vector-meson exchange\label{sec-vm}\label{sec-lec}\label{sec-tpe}}

The one-vector exchange, graph (c), is $V^{(c)}=V_\rho+V_\omega$ with
\be
V_{\rho}&=&
\bar u_1(\vp\,')\,\Gamma_{\rho NN}^{\mu,a}(q)\,u_1(\vp)\,
D_\rho(t)
\no\\ && \times\,
\bar u_2(-\vp\,')\,\Gamma_{\rho NN,\mu}^a(-q)\,u_2(-\vp),
\label{Vborn}\ee
where $u_{1,2}$ are Dirac spinors, $q^\mu=(p'-p)^\mu$, and $\vp\,'$ and
$\vp$ are the relative momenta in the center-of-mass system.
$V_\omega$ can be obtained by replacing $\Gamma_{\rho NN}^{\mu,a}$
by $\Gamma_{\omega NN}^{\mu}$
and $D_\rho$ by $D_\omega$.

The $\rho NN$ and $\omega NN$ vertex functions can be written as
\be
\Gamma_{\rho NN}^{\mu,a}(q) &=& g \frac{\tau^a}{2}
\left[\gamma^\mu \Gamma_{1\rho}(t)
+ \frac{i}{2m_N}\Gamma_{2\rho}(t)\sigma^{\mu\nu}q_\nu\right],
\no\\
\Gamma_{\omega NN}^{\mu}(q) &=& g_\omega
\left[\gamma^\mu \Gamma_{1\omega}(t)
+ \frac{i}{2m_N}\Gamma_{2\omega}(t)\sigma^{\mu\nu}q_\nu\right],
\label{Gvertex}\ee
with $t=q^2$.  The Sachs
combinations are
\be
\Gamma_{E x}(t) &=& \Gamma_{1x}(t)+\frac{t}{4m_N^2}\Gamma_{2x}(t),
\no\\
\Gamma_{Mx}(t) &=& \Gamma_{1x}(t)+\Gamma_{2x}(t),
\qquad
x=\rho,\,\omega .
\label{sachs}\ee
The explicit forms of the vertex functions up to ${\cal O}(Q^3)$ are given in Appendix~\ref{app-vert}.

Expanded in $1/m_N$ with the normalization convention $\bar uu=1$~\cite{kgw1998},
Eq.~(\ref{Vborn}) becomes
\be
V_\rho &=& \tt\,\frac{g^2}{4}D_\rho(t)
\Bigg[\Gamma_{E\rho}^2(t)\left(1+\frac{2k^2}{m_N^2}\right)
\no\\ &&
+\,\frac{\left(\sqq-q^2\ssig\right) \Gamma_{M\rho}^2(t)}{4m_N^2}
\no\\ &&
-\,\frac{3\Gamma_{1\rho}^2(t)+4\Gamma_{1\rho}(t)\Gamma_{2\rho}(t)}{2m_N^2}\,\OLS\Bigg],
\label{VVM}\ee
where $\vk=(\vp\,'+\vp)/2$ and
$\OLS=-\frac{i}{2}(\vsigma_1+\vsigma_2)\!\cdot\!(\vq\times\vk)$.
The $k^2$ term is the $1/m_N^2$ correction to the central term.
Terms of order $1/m_N^4$ are omitted.
The
$\omega$ contribution follows on replacing $\tt\,g^2/4$ by $g_\omega^2$
and $\rho$ by $\omega$.

The potentials of graphs (a) and (b) can be computed directly,
but there is a shorter route once the ChPT potential is known.
Let $V_{\rm VM}^{(a,b,c)}$ be the potentials of the theory with vector
mesons and $V_{\rm ChPT}^{(a,b)}$ those of ChPT.
Let $V_{\rm VM}^{(c,{\rm trunc})}$ be the potential obtained by expanding the propagator and vertex functions in $V_{\rm VM}^{(c)}$ and truncating at
the order considered, here N$^2$LO.  Summed over the graphs, the potential
with vector mesons must then coincide with that of ChPT,
$V_{\rm VM}^{(a)}+V_{\rm VM}^{(b)}+V_{\rm VM}^{(c,{\rm trunc})}
=V_{\rm ChPT}^{(a)}+V_{\rm ChPT}^{(b)}$.  Since
$V_{\rm VM}^{(a)}=V_{\rm ChPT}^{(a)}$,
\be
V_{\rm VM}^{(b)}= V_{\rm ChPT}^{(b)}- V_{\rm VM}^{(c,{\rm trunc})} .
\label{Vb}\ee
Up to NLO, $V_{\rm VM}^{(b)}$ vanishes.
For $a_N=2$ the Weinberg--Tomozawa vertex is replaced by $\rho$ exchange.  
At N$^2$LO, $c_4^{\rm ChPT}+1/(4m_N)$ in $W_T$ is replaced by $c_4^{\rm VM}$,
the $\rho$ exchange carrying the remaining $(1+\kappa_\rho)/(4m_N)$,
consistent with Eq.~(\ref{c4split}).

As discussed in Sec.~II\,A, replacing $V_{\rm VM}^{(c)}$ by 
$V_{\rm VM}^{(c,{\rm trunc})}$
reduces the result to ChPT.
The effect of the explicit vector mesons therefore lies in the difference between N$^2$LO$_{\rm VM}$ and N$^2$LO,
which is of higher order than N$^2$LO in the counting of ChPT.
The full $V_{\rm VM}^{(c)}$ is retained,
with the self-energy and vertex functions evaluated to the corresponding orders,
as discussed in Sec.~\ref{sec-prop}.
The same dressed propagator and vertex functions are used in all terms,
including those that already carry $1/m_N^2$.

The subleading $\pi N$ constants are taken from the Roy--Steiner analysis of
Refs.~\cite{rs-matching,rs-report}: $c_1=-0.74(2)$, $c_3=-3.61(5)$ and
$c_4=2.17(3)$~GeV$^{-1}$.\footnote{We use the value of Ref.~\cite{rs-matching} from the row that pairs with an N$^2$LO potential,
and keep the recoil term
$1/(4m_N)$ explicit in Eq.~(\ref{WTn2lo}).  The $NN$-counting value
$2.44(3)$~GeV$^{-1}$~\cite{emn2017,sms2018,rmn-review} corresponds to
$c_4+1/(4m_N)$.}

Charge-symmetry-breaking $\rho$--$\omega$ mixing~\cite{ge-svenne} is omitted in this work.

\subsection{Regularization\label{sec-reg}}

The spectral-function regularization (SFR) of Ref.~\cite{egm-sfr} removes the two-pion-exchange contributions above the mass $\tilde\Lambda$ in the spectral representation.
In the SFR,
the loop functions
$L(q)$ and $A(q)$ of Eq.~(\ref{LA})
become
\be
L(q;\tilde \Lambda) &=&
(4m_\pi^2+q^2)
\int_{2m_\pi}^{\tilde\Lambda}\!
\frac{d\mu}{\sqrt{\mu^2-4m_\pi^2}\,(\mu^2+q^2)}
\no\\
&=& \frac{\sqrt{4m_\pi^2+q^2}}{q}
\ln\frac{\tilde\Lambda\sqrt{4m_\pi^2+q^2}+q\sqrt{\tilde\Lambda^2-4m_\pi^2}}
{2m_\pi\sqrt{\tilde\Lambda^2+q^2}} ,
\no\\
A(q; \tilde \Lambda) &=& \frac12\int_{2m_\pi}^{\tilde\Lambda}\!\frac{d\mu}{\mu^2+q^2}
\no\\
&=& \frac{1}{2q}\arctan\frac{q(\tilde\Lambda-2m_\pi)}{q^2+2\tilde\Lambda m_\pi} .
\label{LAsfr}\ee
Ref.~\cite{egm-sfr} takes
$\tilde\Lambda$ between $0.5$ and $0.8$~GeV.

Applying the SFR to the vector meson sector
is however complicated for the following reason.
In Ref.~\cite{paper1},
the LECs of the $\rho$ sector are fixed
at the $\rho$ mass shell.
This prescription cannot be used for $\tilde\Lambda<m_\rho$, where the
$\rho$ self-energy loses its imaginary part at $t=m_\rho^2$.
When the cutoff is above the $\rho$ mass, such a complication does not occur.

In this work, we explore three cases of the cutoff
scheme.
The first case is to apply the cutoff only to the two-pion exchange,
graphs (a) and (b) of Fig.~\ref{fig-tpe}, with $\tilde\Lambda$ between $0.5$
and $1.0$~GeV (Table~\ref{tab-rob}), while the $\rho$ self-energy and the
vertex functions are not regularized.
The second case is to apply $\tilde\Lambda=1.0$~GeV to both the two-pion-exchange part and the vector meson
part.
In this case, the LECs of the $\rho$ sector are refitted from the same inputs,
the $\rho\to\pi\pi$ width and the position and
residue of the propagator pole, and the $\rho NN$ vertex slopes are refitted to the nucleon radii.
The third case is
the unregularized
limit $\tilde\Lambda\to\infty$.

The semilocal momentum-space regulator of Ref.~\cite{sms2018} multiplies the exchange of a meson of mass $m$ by
$\exp[-(q^2+m^2)/\Lambda^2]$.
While this
preserves the pole contribution, it strongly suppresses the vector-meson contribution,
even at small $q$, unless $\Lambda$ is well above $m_V$.
Near $q=0$ this factor for $\rho$
exchange is $0.14$ ($0.29$) at $\Lambda=0.55$~GeV ($0.7$~GeV).
For this reason, this scheme is not adopted here.

\begin{table*}
\centering
\caption{The $J=3$ waves at $T_{\rm lab}=150$ and $300$~MeV, in degrees.
The first row is N$^2$LO of ChPT.  The second row is the pion part of
N$^2$LO$_{\rm VM}$, graphs (a1)--(b3) of Fig.~\ref{fig-tpe}.
The next four rows are the Born phase
shifts of each term of Eq.~(\ref{VVM}).
The $\omega$ row is the Born phase shift of the full
$\omega$ exchange.  
\label{tab-j3}}
\begin{tabular}{l@{\ \ }r@{\ \ }r@{\ \ }r@{\ \ }r@{\ \ }r@{\qquad}r@{\ \ }r@{\ \ }r@{\ \ }r@{\ \ }r}
\hline\hline
 & \multicolumn{5}{c}{$150$~MeV} & \multicolumn{5}{c}{$300$~MeV}\\
 & $\nucst{1}{F}{3}$ & $\nucst{3}{D}{3}$ & $\nucst{3}{F}{3}$ & $\nucst{3}{G}{3}$ & $\epsilon_3$
 & $\nucst{1}{F}{3}$ & $\nucst{3}{D}{3}$ & $\nucst{3}{F}{3}$ & $\nucst{3}{G}{3}$ & $\epsilon_3$\\
\hline
N$^2$LO (ChPT)                 & $-2.90$ & $2.82$  & $-1.83$ & $-1.83$ & $4.90$  & $-3.70$ & $8.62$ & $-2.18$ & $-4.63$ & $7.52$\\
\hline
$(1\pi+2\pi)_{\rm VM}$         & $-2.85$ & $2.10$  & $-1.82$ & $-1.85$ & $4.97$  & $-3.48$ & $6.78$  & $-2.15$ & $-4.75$ & $7.82$\\
$\rho$, $\Gamma_{E\rho}^2$     & $0.16$  & $0.82$  & $-0.05$ & $0.03$  & $0$     & $0.55$  & $1.86$  & $-0.18$ & $0.15$  & $0$\\
$\rho$, $k^2$ & $0.04$ & $0.18$ & $-0.01$ & $0.01$  & $0$     & $0.23$  & $0.56$ & $-0.08$ & $0.07$  & $0$\\
$\rho$, $\Gamma_{M\rho}^2$     & $-0.33$ & $0.57$  & $0.03$  & $0.04$  & $-0.20$ & $-1.35$ & $-0.01$ & $0.13$  & $0.23$  & $-0.87$\\
$\rho$, $\OLS$                 & $0$     & $-0.51$ & $-0.02$ & $0.04$  & $0$     & $0$     & $0.03$  & $-0.11$ & $0.27$  & $0$\\
$\omega$                       & $-0.12$ & $-1.31$ & $-0.16$ & $-0.02$ & $0.02$  & $-0.79$ & $-4.82$ & $-1.08$ & $-0.24$ & $0.12$\\
\hline
N$^2$LO$_{\rm VM}$             & $-3.09$ & $1.86$  & $-2.04$ & $-1.75$ & $4.79$  & $-4.84$ & $4.38$  & $-3.46$ & $-4.20$ & $7.13$\\
\hline
PWA93                          & $-2.70$ & $2.74$  & $-1.98$ & $-1.74$ & $4.80$  & $-4.72$ & $4.62$  & $-2.73$ & $-3.90$ & $6.80$\\
\hline\hline
\end{tabular}
\end{table*}

\section{Results\label{sec-results}}

\subsection{Peripheral waves\label{sec-periph}\label{sec-j3}}

\begin{figure*}
\centering
\includegraphics[width=0.93\textwidth]{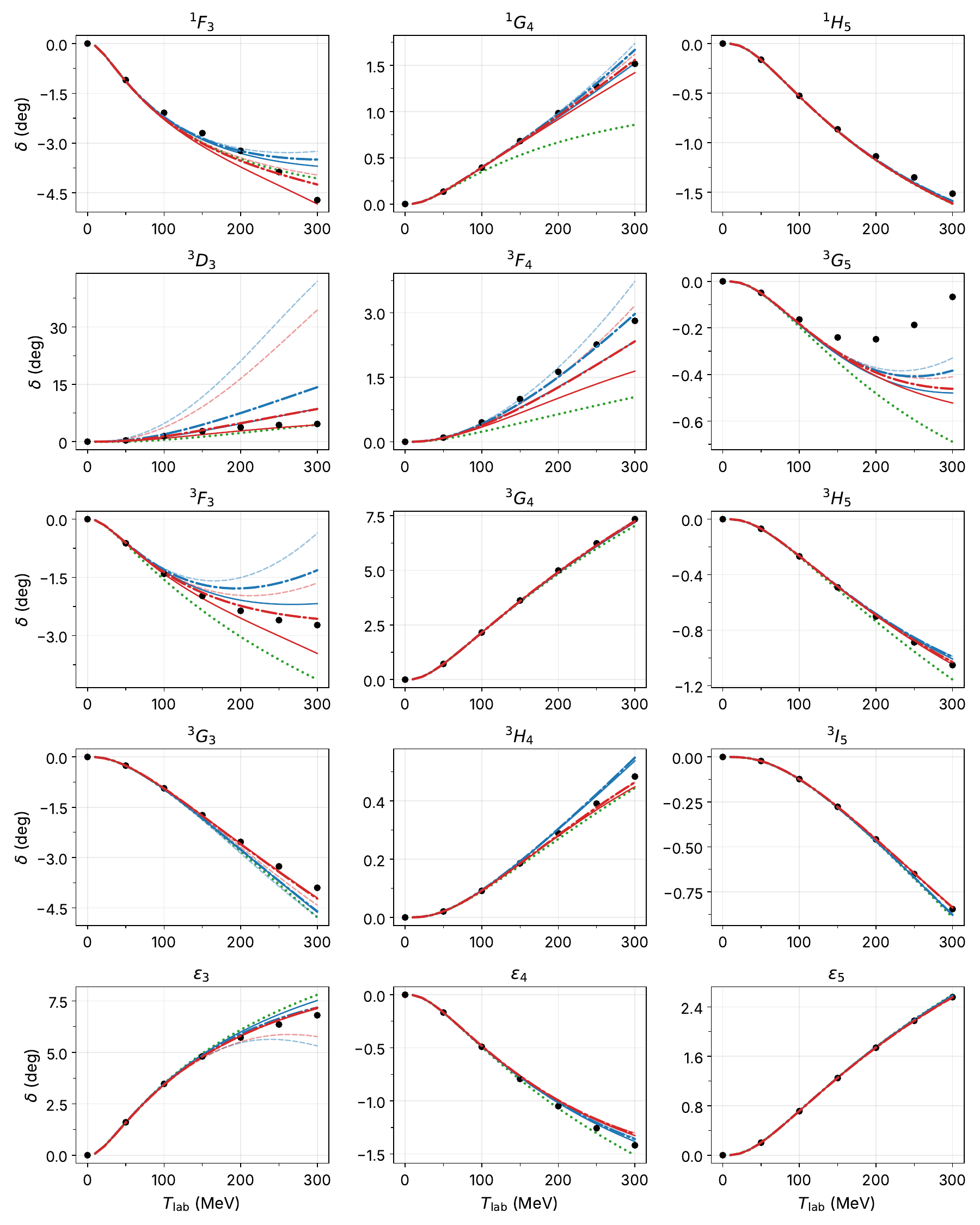}
\caption{Peripheral $np$ phase shifts, one column per total angular momentum.
Green dotted is the common LO long-range result, blue is N$^2$LO of ChPT,
and red is N$^2$LO$_{\rm VM}$, with the matching of Sec.~\ref{sec-lec}.
Solid curves use the spectral cutoff $\tilde\Lambda=0.7$~GeV.
Dot-dashed curves use $\tilde\Lambda=1.0$~GeV applied to
the two-pion exchange and, for N$^2$LO$_{\rm VM}$, to the two-pion loops
of the $\rho$ self-energy and of the $\rho NN$ vertex functions as well,
with the $\rho$-sector constants and the vertex slopes redetermined under the same cut (Sec.~\ref{sec-cutoff}).
Dashed curves use $\tilde\Lambda=\infty$.
LO carries no two-pion loop and is
drawn once.  Circles are PWA93~\cite{nijmegen}.
\label{fig-periph}}
\end{figure*}

Figure~\ref{fig-periph} shows the $np$ nuclear-bar phase shifts of the peripheral waves with $J=3$--$5$ up to $T_{\rm lab}=300$~MeV.
The green dotted curve is LO, the blue curves are N$^2$LO of ChPT and the red curves are N$^2$LO$_{\rm VM}$.
Solid curves use $\tilde\Lambda=0.7$~GeV,
dot-dashed curves $\tilde\Lambda=1.0$~GeV with the SFR in both sectors (Sec.~\ref{sec-reg}),
and dashed curves $\tilde\Lambda=\infty$.
PWA93~\cite{nijmegen} phase shifts are denoted by filled circles.  
The distance
between the blue and the red curves is the vector-meson effect.  It is
concentrated in $J=3$ and decreases rapidly for $J=4$ and 5.

Table~\ref{tab-j3} gives the contributions of the individual terms of $V_\rho$,
Eq.~(\ref{VVM}), to the $J=3$ waves at 150 and 300~MeV,
each in the Born approximation.
The pion part of N$^2$LO$_{\rm VM}$ (row $(1\pi+2\pi)_{\rm VM}$) and the $\omega$ contribution are shown as a whole.

The $\Gamma_{M\rho}^2$ term carries the tensor and spin-spin interaction of
the $\rho$.  
Its role can be seen 
in $\epsilon_3$, to which 
only this term of $V_\rho$ contributes.  At 300~MeV it lowers the mixing
angle by $0.87^\circ$ and works against one-pion exchange.  In the spin
singlet $\nucst{1}{F}{3}$ this term acts strongly through its spin-spin
part, partly cancelled by the $\Gamma_{E\rho}^2$ term.  In
$\nucst{3}{G}{3}$, by contrast, the $\rho$ shift is shared among the
$\Gamma_{M\rho}^2$, $\OLS$ and $\Gamma_{E\rho}^2$ terms.

$\nucst{3}{G}{3}$ is one of the channels that show the vector-meson effect clearly.
The vector mesons
move it toward PWA93 over the whole energy range, and the gap to PWA93 is reduced from
$0.73^\circ$ to $0.30^\circ$ at 300~MeV ($0.09^\circ$ to $0.01^\circ$ at 150~MeV).
The shift is carried mainly by the $\rho$, and it depends little on the spectral
cutoff, on whether the SFR is applied to the vector sector as well, or on
the treatment of iterated OPE (Secs.~\ref{sec-iter} and
\ref{sec-cutoff}).  The same holds, on a much smaller scale, for the other
$L=J+1$ triplet waves $\nucst{3}{H}{4}$ and $\nucst{3}{I}{5}$, where the
small shifts bring the phases onto PWA93.

$\epsilon_3$, which is sensitive to the tensor part of the $\rho$,
improves as well, the gap at 300~MeV falling from $0.72^\circ$ to $0.33^\circ$.

In $\nucst{3}{F}{3}$ the vector mesons add repulsion and carry the phase past PWA93.
At 300~MeV, N$^2$LO$_{\rm VM}$ lies $0.73^\circ$ below PWA93 (N$^2$LO $0.55^\circ$ above).
This wave depends strongly on the spectral cutoff (Sec.~\ref{sec-cutoff}).

In $\nucst{3}{D}{3}$ the $\omega$ contribution,
$-4.8^\circ$ at 300~MeV, outweighs the $\rho$ contribution of $2.4^\circ$.
It comes from the central repulsion, $-4.2^\circ$ from the $\Gamma_{E\omega}^2$ term and $-2.3^\circ$ from the $k^2$ term,
partly offset by the spin-orbit term, $2.0^\circ$.
Together with the pion part of N$^2$LO$_{\rm VM}$, the gap 
at 300~MeV falls from $4.0^\circ$ to $0.2^\circ$.
The pion-exchange part is strongly cutoff dependent in this wave,
and the wave is very sensitive to the iteration of the potential (Sec.~\ref{sec-iter}),
as also observed in Refs.~\cite{egm-sfr,lg2026}.

In $\nucst{1}{F}{3}$ the vector mesons add repulsion at all energies.
At 300~MeV this 
reduces the gap from $1.02^\circ$ to $0.12^\circ$.
Below about 200~MeV, where N$^2$LO is already close to PWA93, it widens the gap.

In $\nucst{3}{F}{4}$ the ChPT curve falls short of PWA93 and the vector mesons move it further away,
widening the gap at 300~MeV from $0.48^\circ$ to $1.17^\circ$.
Ref.~\cite{ekmn2015} finds substantial N$^3$LO and N$^4$LO
two-pion-exchange corrections in this channel.

Of the remaining $J=4$ and 5 waves, $\nucst{3}{G}{4}$, $\nucst{3}{H}{5}$ and $\epsilon_5$ are
reproduced by both calculations, and in $\nucst{1}{G}{4}$, $\epsilon_4$
and $\nucst{1}{H}{5}$ both agree with PWA93 to within about $0.1^\circ$,
N$^2$LO slightly better.
\subsection{Dependence on OPE iteration\label{sec-iter}}

In this calculation, we adopt the peripheral prescription in which
OPE is iterated once, while the remaining interaction is treated at first
order~\cite{kbw,kgw1998,ekmn2015,rmn-review}.  In Table~\ref{tab-iter} we
compare this prescription with OPE in Born approximation and with fully
iterated OPE.  Discussions of perturbative treatments of OPE in peripheral
waves can be found, for example, in Refs.~\cite{pvsyl2017,kaplan2020}.

\begin{table}
\centering
\caption{Dependence of N$^2$LO$_{\rm VM}$ on the treatment of OPE,
taken in Born approximation, once iterated or fully iterated,
with the rest of the potential at first order.  
Phase shifts are in degrees.
\label{tab-iter}}
\begin{tabular}{l@{\ }r@{\ }r@{\ }r@{\ }r@{\ \ \ }r@{\ }r@{\ }r@{\ }r}
\hline\hline
 & \multicolumn{4}{c}{$150$~MeV} & \multicolumn{4}{c}{$300$~MeV}\\
 & $\nucst{3}{D}{3}$ & $\nucst{3}{F}{3}$ & $\nucst{3}{G}{3}$ & $\epsilon_3$
 & $\nucst{3}{D}{3}$ & $\nucst{3}{F}{3}$ & $\nucst{3}{G}{3}$ & $\epsilon_3$\\
\hline
Born      & $-2.45$ & $-2.11$ & $-1.38$ & $5.09$ & $-5.76$ & $-3.63$ & $-2.63$ & $7.82$\\
Once      & $1.86$ & $-2.04$ & $-1.75$ & $4.79$ & $4.38$ & $-3.46$ & $-4.20$ & $7.13$\\
Full      & $1.39$ & $-2.04$ & $-1.73$ & $4.79$ & $3.05$ & $-3.48$ & $-4.06$ & $6.82$\\
\hline
PWA93     & $2.74$ & $-1.98$ & $-1.74$ & $4.80$ & $4.62$ & $-2.73$ & $-3.90$ & $6.80$\\
\hline\hline
\end{tabular}
\end{table}

$\nucst{3}{D}{3}$ depends very strongly on the treatment of OPE at both
energies, and $\nucst{3}{G}{3}$ and $\epsilon_3$ also show a sizable
dependence at 300~MeV.  The difference between the N$^2$LO$_{\rm VM}$ and
N$^2$LO results in $\nucst{3}{G}{3}$, however, changes by only $0.01^\circ$
among the three prescriptions.  The $J=4$
and 5 waves have a much milder dependence.  The once-iterated and fully
iterated prescriptions agree to within $0.1^\circ$.

\subsection{Dependence on the spectral cutoff and on the vertex functions
\label{sec-cutoff}\label{sec-cov}}

The dependence of N$^2$LO$_{\rm VM}$ on the spectral cutoff $\tilde\Lambda$ is similar to that of N$^2$LO (Fig.~\ref{fig-periph}) and is listed in Table~\ref{tab-rob} for the $J=3$ triplet waves.
The rows $\tilde\Lambda=0.5$, 0.7 and 1.0~GeV correspond to the first case of Sec.~\ref{sec-reg},
where the cutoff is applied only to the two-pion exchange,
and the row $1.0^*$ to the second, where it is applied to the vector sector as well.
The row $\infty$ is the unregularized limit.

\begin{table}[!h]
\centering
\caption{Dependence of N$^2$LO$_{\rm VM}$ on the spectral cutoff $\tilde\Lambda$ of two-pion exchange,
with the vector-sector input kept fixed.
In the row $1.0^*$ the SFR is applied to the vector sector as well.
Phase shifts are in degrees.
\label{tab-rob}}
\begin{tabular}{l@{\ }r@{\ }r@{\ }r@{\ }r@{\ \ \ }r@{\ }r@{\ }r@{\ }r}
\hline\hline
 & \multicolumn{4}{c}{$150$~MeV} & \multicolumn{4}{c}{$300$~MeV}\\
$\tilde\Lambda$~[GeV] & $\nucst{3}{D}{3}$ & $\nucst{3}{F}{3}$ & $\nucst{3}{G}{3}$ & $\epsilon_3$
 & $\nucst{3}{D}{3}$ & $\nucst{3}{F}{3}$ & $\nucst{3}{G}{3}$ & $\epsilon_3$\\
\hline
$0.5$     & $1.15$ & $-2.24$ & $-1.75$ & $4.79$ & $2.33$ & $-4.40$ & $-4.26$ & $7.16$\\
$0.7$     & $1.86$ & $-2.04$ & $-1.75$ & $4.79$ & $4.38$ & $-3.46$ & $-4.20$ & $7.13$\\
$1.0$     & $3.04$ & $-1.90$ & $-1.75$ & $4.77$ & $8.61$ & $-2.60$ & $-4.20$ & $7.01$\\
$1.0^*$   & $3.00$ & $-1.89$ & $-1.76$ & $4.79$ & $8.54$ & $-2.57$ & $-4.23$ & $7.17$\\
$\infty$  & $9.10$ & $-1.79$ & $-1.79$ & $4.71$ & $34.42$ & $-1.65$ & $-4.42$ & $5.77$\\
\hline
PWA93     & $2.74$ & $-1.98$ & $-1.74$ & $4.80$ & $4.62$ & $-2.73$ & $-3.90$ & $6.80$\\
\hline\hline
\end{tabular}
\end{table}

$\nucst{3}{G}{3}$ and $\epsilon_3$ depend weakly on the cutoff.
At 300~MeV N$^2$LO$_{\rm VM}$ stays $0.3^\circ$--$0.5^\circ$ below PWA93 in $\nucst{3}{G}{3}$ and gives a better description than N$^2$LO at every cutoff.
$\epsilon_3$ changes little up to $\tilde\Lambda=1$~GeV and drops in the unregularized limit, as N$^2$LO does.
$\nucst{3}{D}{3}$ follows the strong cutoff dependence of the pion-exchange part.
$\nucst{3}{H}{4}$ and $\nucst{3}{I}{5}$ are likewise weakly cutoff dependent.

The rows 1.0 and $1.0^*$ of Table~\ref{tab-rob} are close to each other,
so the first and second cases of the cutoff scheme discussed in Sec.~\ref{sec-reg} give nearly the same results.\footnote{Applying the SFR to the vector sector at $\tilde\Lambda=1.0$~GeV moves $g$ from $6.21$ to $6.45$ and $A_\rho(0)$ from $1.11$ to $1.26$,
while the pole remains at $m_\rho$.}

The heavy-baryon expansion used for the $\rho NN$ vertex loops is known to converge slowly for the nucleon electromagnetic form factors,
and covariant evaluations of these loops are adopted in Refs.~\cite{kubis2001,fgs2004,schindler2005}.
Repeating the calculation with the covariant loops changes the N$^2$LO$_{\rm VM}$ phases at $\tilde\Lambda=0.7$~GeV by at most $0.13^\circ$ in $\nucst{3}{G}{3}$,
$0.05^\circ$ in $\epsilon_3$ and $0.21^\circ$ in $\nucst{3}{D}{3}$.
The conclusions are therefore unchanged with the covariant vertex,
although the improvement over N$^2$LO in $\nucst{3}{G}{3}$ is reduced by about 30\%.

\subsection{Dependence on $\kappa_\rho$,
$\kappa_\omega$, $g_\omega$ and $c_4^{\rm VM}$\label{sec-input}}

We examine how the peripheral phase shifts depend on $\kappa_\rho$,
$\kappa_\omega$, $g_\omega$ and $c_4^{\rm VM}$.
These constants are varied such that the experimental nucleon magnetic moments and electromagnetic radii are reproduced.
Varying $\kappa_\omega$ within its phenomenological range~\cite{mmd1996,zhu1999,cdbonn} has little effect,
while $\kappa_\rho$ has a more noticeable influence.
Fitting $\kappa_\rho$, $\kappa_\omega$ and $c_4^{\rm VM}$ simultaneously to PWA93 yields only a small improvement at the cost of large deviations from their reference values.
Increasing $g_\omega$ toward the values used in Refs.~\cite{kgw1998,cdbonn} worsens the agreement,
mainly in $\nucst{3}{D}{3}$ and $\nucst{3}{F}{3}$.  

\section{Summary}

We have studied peripheral $NN$ scattering in ChPT with the $\rho$ and $\omega$ kept as explicit degrees of freedom.
The key aspect of the approach is to count their momenta as of order $Q$ and their masses as of order $\Lambda_\chi$.
The vector propagators and $VNN$ vertices are taken from the mesonic sector and the nucleon electromagnetic form factors.
The calculation is performed up to N$^2$LO,
where part of the two-pion exchange of ChPT is replaced by $\rho$ exchange.  
Our prediction, N$^2$LO$_{\rm VM}$,
differs from N$^2$LO only by the higher-order terms resummed in the vector-meson propagators and vertices.

Among the waves considered, the vector-meson effect is concentrated in the $J=3$ waves.
The $\rho$ moves $\nucst{3}{G}{3}$ and $\epsilon_3$ toward PWA93.
The improvement in $\nucst{3}{G}{3}$ persists over the whole range of the spectral cutoff,
and the shift is nearly independent of the treatment of OPE.
The $\omega$ adds central repulsion, which reduces the gap to PWA93 in $\nucst{3}{D}{3}$ at 300~MeV and widens it in $\nucst{3}{F}{4}$.
$\nucst{3}{D}{3}$ remains strongly dependent on the treatment of pion exchange.

Applying the spectral cutoff to the vector sector as well,
with a cutoff above the $\rho$ mass, has little effect on the results.
Covariant vertex loops in place of the heavy-baryon ones leave these features unchanged.

\begin{acknowledgments}
This work was supported by the IBS grant funded by the Korea government
(No. IBS-R031-D1).
\end{acknowledgments}

\section*{Data Availability}
The data are not publicly available.
The data are available from the author upon reasonable request.

\appendix

\section{The vector-meson--nucleon vertex functions\label{app-vert}}

The $\rho NN$ vertex functions up to ${\cal O}(Q^3)$ read
\be
\Gamma_{1\rho}(t) &=& 1 + \Gamma_{1\rho}'(0)\,t + L_E(t) ,
\no\\
\Gamma_{2\rho}(t) &=& \kappa_\rho + \Gamma_{2\rho}'(0)\,t + L_M(t) ,
\no\\
L_E(t) &=& -\frac{g_A^2\,t}{16\pi^2f_\pi^2}\left[\frac32 K_0(t)-4K_2(t)\right] ,
\no\\
L_M(t) &=& -\frac{g_A^2m_N}{8\pi f_\pi^2}\left[(4m_\pi^2-t)
A\left(\sqrt{-t}\right)-m_\pi
+\frac{t}{6m_\pi}\right]
\no\\ &&
+\ \frac{c_4^{\rm VM}\,m_N\,t}{8\pi^2 f_\pi^2}
\left[4K_2(t)-K_0(t)
\right]
\label{vertloop}\ee
with $A(q)$ of Eq.~(\ref{LA}) and
\be
K_0(t) &=& \int_0^1\!dz\,\ln\!\left[1-\frac{z(1-z)\,t}{m_\pi^2}\right] ,
\no\\
K_2(t) &=& \int_0^1\!dz\,z(1-z)\,
\ln\!\left[1-\frac{z(1-z)\,t}{m_\pi^2}\right] .
\label{Kfns}\ee
In terms of the two-pion loop functions
\be
L(q) &=& \frac{w}{q}\ln\frac{w+q}{2m_\pi} ,
\no\\
A(q) &=& \frac{1}{2q}\arctan\frac{q}{2m_\pi} ,
\label{LA}\ee
with $q=\sqrt{-t}$ and $w=\sqrt{4m_\pi^2+q^2}$,
$K_0(t)=2L(q)-2$ and
$K_2(t)=\frac1{18}+\frac{q^2-2m_\pi^2}{3q^2}\left[L(q)-1\right]$.
The isoscalar vertex carries no loop at this order,
\be
\Gamma_{1\omega}(t) = 1 + \Gamma_{1\omega}'(0)\,t ,
\qquad
\Gamma_{2\omega}(t) = \kappa_\omega + \Gamma_{2\omega}'(0)\,t .
\label{vertomega}\ee
The loop functions are defined such that $L_{E,M}(t)={\cal O}(t^2)$, and
the slopes of the vertex functions at $t=0$ read~\cite{bkkm1992,kubis2001}
\be
\Gamma_{1\rho}'(0) &=& -2d_6^\rho(\lambda)
-\frac{5g_A^2}{96\pi^2f_\pi^2}\ln\frac{m_\pi^2}{\lambda^2} ,
\no\\
\Gamma_{2\rho}'(0) &=& 4m_N e_{74}^\rho + \frac{g_A^2m_N}{48\pi f_\pi^2m_\pi} ,
\no\\
\Gamma_{1\omega}'(0) &=& -4d_7^\omega ,
\qquad
\Gamma_{2\omega}'(0) = 8m_N e_{54}^\omega ,
\label{lecslopes}\ee
where $\lambda$ is the renormalization scale, and the $d$ terms have
$\nu_i=2$ and the $e$ terms $\nu_i=3$ in the counting of
Eq.~(\ref{counting}).

The LECs are determined from the relation of the vertex functions to
the nucleon electromagnetic form factors,
\be
G_{X}^V(t) &=& \hat D_\rho(t)\,\Gamma_{X\rho}(t) ,
\no\\
G_{X}^S(t) &=& \hat D_\omega(t)\,\Gamma_{X\omega}(t) ,
\no\\
\hat D_\rho(t) &=& \left(1+\beta_\rho t\right) 2g^2f_\pi^2 D_\rho(t) ,
\no\\
\hat D_\omega(t) &=& \frac{m_\omega^2}{m_\omega^2-t} ,
\label{GVS}\ee
with $G_X^{S,V}(t)=G_X^p(t)\pm G_X^n(t)$ for $X=E,M$.  Here
$\beta_\rho=C_{1:3}/(2f_\pi^2)=0.250$~GeV$^{-2}$, with $C_{1:3}$ the LEC introduced
in Ref.~\cite{paper1}.

The LECs $d_6^\rho$, $d_7^\omega$, $e_{74}^\rho$ and $e_{54}^\omega$
are fixed by the electric and magnetic radii,
${G_X^{S,V}}'(0)/G_X^{S,V}(0)=r_{XS,XV}^2/6$ with $r_{ES,EV}^2=r_{Ep}^2\pm r_{En}^2$ and $G_M^{S,V}(0)\,r_{MS,MV}^2=\mu_p r_{Mp}^2\pm\mu_n r_{Mn}^2$.
The slopes then follow as
\be
\frac{\Gamma_{X\rho}'(0)}{\Gamma_{X\rho}(0)} &=& \frac{r_{XV}^2}{6}
- \hat D_\rho'(0),
\nonumber\\
\frac{\Gamma_{X\omega}'(0)}{\Gamma_{X\omega}(0)}  &=& \frac{r_{XS}^2}{6}
- \frac{1}{m_\omega^2},
\label{slopes}\ee
for $X=E,M$, where $\hat D_\rho'(0)=\beta_\rho+A_\rho(0)/(2g^2f_\pi^2)$, and
Eq.~(\ref{sachs}) turns the Sachs slopes into $\Gamma_1'(0)$ and
$\Gamma_2'(0)$.
We use the CODATA proton
charge radius~\cite{codata2022}, $r_{Ep}^2=0.707$~fm$^2$, the neutron charge
radius~\cite{pdg}, $r_{En}^2=-0.116$~fm$^2$, and the
magnetic radii of Ref.~\cite{lhm2022}, $r_{Mp}^2=0.721$ and
$r_{Mn}^2=0.746$~fm$^2$.
This results in
$\Gamma_{1\rho}'(0)=0.517$, $\Gamma_{2\rho}'(0)=5.025$,
$\Gamma_{1\omega}'(0)=0.934$ and $\Gamma_{2\omega}'(0)=0.135$~GeV$^{-2}$.

\section{Two-pion exchange contributions\label{app-tpe}}

The ChPT two-pion-exchange potentials are taken from Ref.~\cite{kbw}.
Of the $1/m_N$ corrections only the recoil term in $W_T$ is kept.
At
NLO,
\be
W_C &=& \frac{L(q)}{384\pi^2 f_\pi^4}
\Big[4m_\pi^2(1+4g_A^2-5g_A^4)
\no\\ && \qquad\ \
+\,q^2(1+10g_A^2-23g_A^4)-\frac{48g_A^4m_\pi^4}{w^2}\Big],
\no\\
V_T &=& -\frac{1}{q^2}V_S
=-\frac{3g_A^4}{64\pi^2f_\pi^4}L(q),
\label{VTPE-NLO}\ee
and at N$^2$LO the isoscalar central potential is
\be
V_C &=& \frac{3g_A^2}{16\pi f_\pi^4}
\left[2m_\pi^2(c_3-2c_1)+c_3q^2\right]
\no\\ && \qquad\quad \times\,
(2m_\pi^2+q^2)A(q) ,
\label{VCn2lo}\ee
and the isovector tensor potential at that order is
\be
W_T &=& -\frac{g_A^2}{32\pi f_\pi^4}
\left(c_4+\frac{1}{4m_N}\right)w^2A(q),
\label{WTn2lo}\ee
with the loop functions $L$ and $A$ of Eq.~(\ref{LAsfr}), which reduce
to Eq.~(\ref{LA}) for $\tilde\Lambda\to\infty$.

In N$^2$LO$_{\rm VM}$ the pion part $V_{\rm VM}^{(a)}+V_{\rm VM}^{(b)}$ is
fixed by Eq.~(\ref{Vb}).  The isoscalar potentials are unchanged,
$V_{j,{\rm VM}}^{(a+b)}=V_j$ for $j=C,S,T$.  In the isovector potentials
the $g_A^0$ and $g_A^2$ terms of $W_C$ are removed and $c_4+1/(4m_N)$ in
$W_T$ is replaced by $c_4^{\rm VM}$,
\be
W_{C,{\rm VM}}^{(a+b)} &=& -\frac{g_A^4L(q)}{384\pi^2f_\pi^4}
\left[20m_\pi^2+23q^2+\frac{48m_\pi^4}{w^2}\right],
\no\\
W_{T,{\rm VM}}^{(a+b)} &=& -\frac{g_A^2}{32\pi f_\pi^4}\,c_4^{\rm VM}\,w^2A(q),
\label{TPEvm}\ee
with $W_{S,{\rm VM}}^{(a+b)}=-q^2W_{T,{\rm VM}}^{(a+b)}$.
The
subtracted loop function of Eq.~(\ref{Arho}) is, in the notation of
Ref.~\cite{paper1},
\be
\bar f_3(t) &=& \frac{1}{16\pi^2}\int_{2m_\pi}^{\infty}\!\frac{dM}{3M}
\left(1-\frac{4m_\pi^2}{M^2}\right)^{3/2}
\no\\ && \qquad\qquad\ \times\,
\frac{t}{M^2-t-i0} .
\label{f3bar}\ee

The two-pion-exchange potentials are evaluated with the average pion mass $m_\pi=(m_{\pi^0}+2m_{\pi^\pm})/3=138.039$~MeV~\cite{machleidt-entem,rmn-review}.

\end{document}